%% file: main.tex
\documentclass[conference,9pt]{IEEEtran}

\usepackage{algorithm}
\usepackage{algpseudocode}
\usepackage{amsmath,amssymb,amsfonts}
\usepackage{array}
\usepackage{booktabs}
\usepackage{cite}
\usepackage{colortbl}
\usepackage{float}
\usepackage{graphicx}
\usepackage{listings}
\usepackage{mathrsfs}
\usepackage{multicol}
\usepackage{multirow}
\usepackage{pifont}
\usepackage{siunitx}
\usepackage{soul}
\usepackage{subcaption}
\usepackage{textcomp}

\usepackage{url}
\usepackage{wasysym}
\usepackage{xcolor}

\usepackage[hidelinks]{hyperref}

\IEEEoverridecommandlockouts

\def\BibTeX{{\rm B\kern-.05em{\sc i\kern-.025em b}\kern-.08em
    T\kern-.1667em\lower.7ex\hbox{E}\kern-.125emX}}

\definecolor{blue}{HTML}{56B4E9}
\definecolor{orange}{HTML}{D55E00}
\definecolor{green}{HTML}{009E73}

\newcommand{\cmark}{\scalebox{1.1}{\textcolor{blue}{\ding{51}}}}
\newcommand{\xmark}{\scalebox{1.1}{\textcolor{orange}{\ding{55}}}}
\newcommand{\pmark}{\scalebox{1.1}{\textcolor{green}{\rotatebox{90}{\LEFTcircle}}}}
\newcommand{\cmarkbold}{\scalebox{1.3}{\textcolor{blue}{\ding{52}}}}

\renewcommand{\baselinestretch}{0.98}

\begin{document}

\title{\vspace{-0.3cm}
DiffPower: GPU-Accelerated Differentiable Switching Power Analysis and Optimization\vspace{-0.3cm}
}


\author{%
Isaac Jacobson$^{1}$, 
Zheng Zhao$^{2}$,
Rashmi Mehrotra$^{2}$,
Guanglei Zhou$^{1}$,
Vineet Rashingkar$^{2}$,
Yiran Chen$^{1}$\\
$^{1}$\textit{Electrical and Computer Engineering, Duke University, Durham, USA}\\
$^{2}$\textit{Synopsys Inc., Sunnyvale, USA}\\
$^{1}$\{isaac.jacobson, guanglei.zhou, yiran.chen\}@duke.edu\\
$^{2}$\{zheng.zhao, rashmi.mehrotra, vineet.rashingkar\}@synopsys.com
\vspace{-0.5cm}
}

\maketitle

\input{tex/0_abstract.tex}


\input{tex/1_introduction.tex}
\input{tex/2_background.tex}

\input{tex/3_method.tex}
\input{tex/4_optimizations}
\input{tex/5_experiment.tex}

\input{tex/6_discussion.tex}
\input{tex/7_conclusion.tex}


\section*{Acknowledgment}
\vspace{-0.1cm}
This work was supported in part by NSF Award 2112562 and a gift from Synopsys.

\bibliographystyle{IEEEtran}
\bibliography{references}

\end{document}

%% file: tex/0_abstract.tex
\begin{abstract}

Accurate and scalable switching power analysis remains a critical bottleneck in modern physical design, often forcing a trade-off between computational speed and modeling fidelity.
We present DiffPower, a GPU-accelerated framework for differentiable power analysis and optimization. DiffPower translates design netlists into a PDK-agnostic bytecode representation, enabling analytical gradient computation via reverse-mode automatic differentiation, achieving up to a $1{,}002\times$ speedup over single-threaded CPU propagation on the largest evaluated design, with the GPU advantage growing with design scale.
A hybrid propagation methodology fusing analytical modeling with parallel simulation achieves a median toggle-rate correlation of $r{=}0.96$ across ten industrial and benchmark designs.
The resulting \emph{power gradients}, computed up to $904\times$ faster than CPU finite-difference methods with near-perfect rank agreement, enable two downstream applications: (1)~gradient-weighted cell sizing, which achieves up to $2.98\times$ improvement over local-power heuristics on industrial designs, with even stronger advantages at the 117K-cell scale where competing methods plateau; and (2)~power virus generation via gradient ascent, which yields up to $2.13\times$ higher transition-weighted power, replacing a search process that traditionally requires hours.

\end{abstract}

%% file: tex/1_introduction.tex
\section{Introduction}

The cessation of Dennard scaling has elevated power efficiency to a primary design objective in modern integrated circuits (ICs)~\cite{esmaeilzadeh2011dark}. In the regime of dark silicon, excessive switching power exacerbates thermal management challenges, constrains the battery life of mobile and edge devices, and imposes severe limits on achievable performance.
As a result, Electronic Design Automation (EDA) flows must prioritize power optimization throughout the design hierarchy, from high-level synthesis to placement and routing.

However, optimizing for switching power introduces a persistent challenge: the ``Analysis-Optimization Gap.''
Precise switching power analysis requires estimating the Toggle Rate (TR) of every net, as dynamic power scales with switching activity and capacitive load.
The industry standard methodology involves vector-based simulations (e.g., VCD analysis) using commercial tools such as Synopsys Fusion Compiler or PrimePower.
While this approach captures temporal correlations to yield high fidelity, its computational latency, which often spans hours or days for multi-million-gate designs, renders it infeasible for integration within a tight, iterative optimization loop. In practice, iterative optimization demands fast approximate models that preserve correct sensitivity rankings rather than exact absolute values. A power model invoked thousands of times per optimization pass need only point the optimizer in the right direction for each step to be productive. This speed-fidelity tradeoff is well-established across EDA, from analytical placers to fast timing estimators.

Traditionally, to bridge this gap without resorting to full simulation, designers frequently rely on Static Probability (SP) propagation methods to analytically estimate switching activity. Despite significantly reduced runtimes, traditional SP methods suffer from three fundamental limitations that severely hinder their adoption in advanced optimization flows:

\begin{enumerate}
    \item \textbf{Latency Bottlenecks:} Serial, level-by-level graph traversal on CPUs remains a throughput bottleneck for multi-million-pin designs, preventing the rapid re-analysis necessary after each incremental netlist change.
    \item \textbf{Sensitivity Blindness:} While traditional probabilistic methods effectively identify \textit{where} power is consumed, they fail to reveal \textit{how} to optimize it. Without the mathematical sensitivity ($\nabla P$, the gradient of total power with respect to local activity), these tools force downstream optimization engines to abandon global mathematical convergence in favor of greedy, localized heuristics.
    \item \textbf{Library and PDK Dependence:} Conventional implementations require per-gate-type code for both SP evaluation and, critically, derivative computation. Supporting a new cell library means implementing forward and backward rules for each unique cell function, making gradient-enabled tools inflexible and expensive to extend across technology generations. General-purpose approaches such as BDD-based propagation avoid this but scale poorly to GPU execution.
\end{enumerate}

To overcome these barriers, we present \textbf{DiffPower}, a high-speed, fully differentiable approximation engine engineered specifically for inner-loop power optimization. Rather than replacing sign-off-level vector simulation, DiffPower provides rapid gradient-guided feedback between commercial tool runs, enabling optimization decisions at scales where traditional methods are infeasible. The advantages are highlighted in Table~\ref{tab:diffpower_comparison}.

\input{tables/1_motivation}

Prior approaches have attempted to address these limitations with varying degrees of success.
Binary Decision Diagrams (BDDs) support exact probability propagation, but grow exponentially in the worst case and do not map well to data-parallel GPU execution~\cite{bryant2018binary}.
Recent GPU-accelerated differentiable EDA frameworks have demonstrated the power of combining hardware parallelism with automatic differentiation: INSTA~\cite{lu2025insta} enables differentiable static timing analysis on a levelized graph, while additional differentiable timing optimizers~\cite{ji2025diffccd, du2025differentiable} target delay and slack.
DiffPower extends this paradigm to switching power, where SP propagation constitutes a continuous relaxation of the Boolean circuit~\cite{petersen2022differentiablelogic}.
For power estimation specifically, GATSPI~\cite{zhang2022gatspigpuacceleratedgatelevel} accelerates gate-level simulation with bit-packed parallel logic on GPUs but does not provide analytical gradients; GRANNITE~\cite{zhang2020grannite} infers toggle rates via a graph neural network but also lacks gradient information.
We adopt GATSPI-style bit-packed simulation to derive per-pin scaling factors that calibrate the differentiable model, so that optimization gradients reflect accurate toggle rates.

To achieve PDK-agnostic gradient computation, we use stack-machine bytecode compiled from Boolean expressions. The same bytecode drives both forward SP propagation and reverse-mode gradient computation. This follows the principle of differentiating an intermediate representation, directly analogous to compiler-level Automatic Differentiation (AD) systems such as Enzyme~\cite{moses2021enzyme}. Compiling to bytecode has precedent in simulation (Icarus Verilog VVP~\cite{williams1998icarus}). DiffPower extends the idea to massively parallel GPU execution, where the zero-address format minimizes warp divergence.

To demonstrate the practical value of power gradients, we apply DiffPower to two downstream tasks where global sensitivity information is critical but currently unavailable: power-aware cell sizing and power virus generation.
Gate sizing is a well-studied NP-hard problem~\cite{ning1994strongly}; most prior work focuses on timing rather than power~\cite{chuang1995timing, lu2025lego, pham2025agd, ye2025learning}. Differentiable methods have also been applied to physical design, such as placement~\cite{lin2019dreamplace} and routing~\cite{li2024dgr}.

Power virus generation seeks to induce near-worst-case switching power in order to stress-test a circuit. Existing state-of-the-art methods rely on evolutionary search~\cite{chatzimiltis2025saga, jiang1998exact, deligiannis2021maximizing, ganesan2010system, deligiannis2023automating}, gate-level cost-benefit heuristics~\cite{hajimiri2015efficient}, or greedy discrete search, all of which scale poorly to large designs. In the broader field of automatic test pattern generation, DEFT~\cite{li2026deftdifferentiableautomatictest} applies differentiable programming to fault detection; DiffPower takes a similar approach to generate power viruses across designs.

Concretely, DiffPower maps the netlist to a levelized pin-level Directed Acyclic Graph (DAG) processed by custom CUDA kernels, enabling end-to-end differentiable power analysis with exact gradients in a single backward pass. It overcomes latency through GPU parallelism, provides sensitivity through automatic differentiation, and achieves library independence through a bytecode abstraction. Our contributions are as follows:
\begin{itemize}
    \item We present \textbf{DiffPower}, the first GPU-accelerated, fully differentiable switching power analysis framework that achieves up to $1{,}002\times$ speedup over CPU propagation, completing forward and backward passes on a 652K-cell design in under 30\,ms each.
    \item We propose a \textbf{PDK-agnostic bytecode representation} that compiles arbitrary Boolean cell functions into stack-machine instructions, enabling exact gradient computation for any cell library without gate-specific derivative code.
    \item We introduce a \textbf{hybrid propagation} methodology that bridges bit-packed simulation with a differentiable model, achieving a median TR correlation of $r{=}0.96$ with an SP~$r > 0.96$ on ten of ten industrial and benchmark designs.
    \item We demonstrate \textbf{gradient-guided cell sizing} that achieves $2.98\times$ improvement over local-power heuristics on an industrial design.
    \item We formulate a highly efficient, \textbf{gradient-ascent-based power virus generation} technique that consistently outperforms evolutionary searches, achieving up to $2.13\times$ higher transition-weighted power.

\end{itemize}

%% file: tables/1_motivation.tex
\begingroup
\addtolength{\textfloatsep}{4pt}
\addtolength{\intextsep}{4pt}
\renewcommand{\arraystretch}{0.93}

\begin{table}[htbp]
\centering
\caption{Comparison of power analysis methodologies and their benefits. Analytical SP is fast compared to Commercial Tools but much slower than DiffPower; it is also accurate for SP but not for TR. \cmark\ = yes, \xmark\ = no, \pmark\ = partial.}
\label{tab:diffpower_comparison}
\begin{tabular}{@{}lccc@{}}
\toprule
\textbf{Feature} & \textbf{Commercial Tools} & \textbf{Analytical SP} & \textbf{DiffPower} \\ \midrule
High Fidelity                & \cmark & \pmark & \cmarkbold \\
Low Latency                  & \xmark & \pmark & \cmarkbold \\
Sensitivity $(\nabla P)$     & \xmark & \xmark & \cmarkbold \\
PDK-Agnostic                 & \xmark & \xmark & \cmarkbold \\ \bottomrule
\end{tabular}
\end{table}
\endgroup

%% file: tex/2_background.tex
\section{Preliminaries} \label{background}

\begin{figure*}[t]
\centering
\includegraphics[width=.95\linewidth]{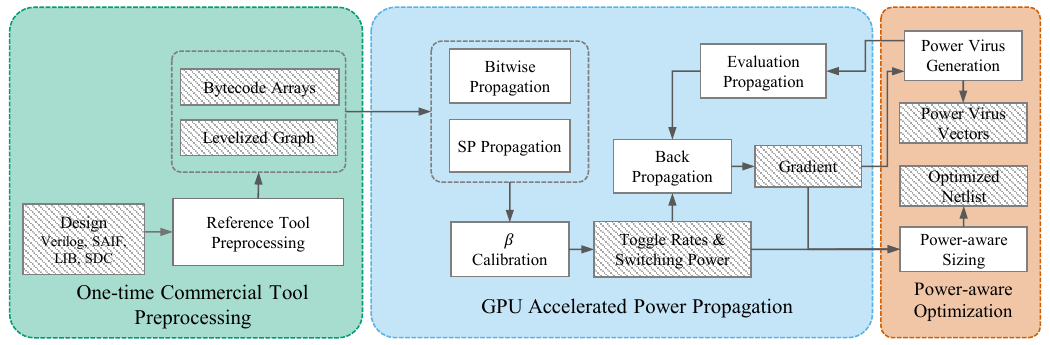}
\caption{Overview of DiffPower. The framework initializes from a commercial reference tool, builds a levelized pin graph with corresponding bytecode arrays loaded to GPU memory (green box on the left). DiffPower propagates static probability values algebraically to model toggle rates, calibrates them via bitwise simulation ($\beta$-calibration), and computes exact gradients through backpropagation (blue box in the middle). These gradients guide cell sizing and power virus generation via gradient ascent (orange box on the right).}
\label{fig:overview}
\end{figure*}

\subsection{Switching Power Model}
The dynamic switching power ($P_{sw}$) of a digital CMOS circuit is formally modeled as the sum of power dissipated by charging and discharging capacitive loads across all nets in the design:
\begin{equation}
    P_{sw} = \frac{1}{2} \cdot V_{dd}^2 \cdot \sum_{\forall i \in \mathrm{Nets}} \alpha_i \cdot f_{clk} \cdot C_{load, i}
\label{eq:psw}
\end{equation}
where $V_{dd}$ denotes the supply voltage, $f_{clk}$ represents the clock frequency, $C_{load, i}$ is the total capacitive load of net $i$ (including wire capacitance and pin capacitance of driven cells), and $\alpha_i$ signifies the switching activity (Toggle Rate) of net $i$. In the context of physical synthesis optimization, $V_{dd}$ and $f_{clk}$ are typically fixed system constraints determined by the operating mode. Therefore, optimization efforts focus on minimizing the sum $\sum \alpha_i \cdot C_{load, i}$ through logic restructuring (affecting $\alpha_i$) and cell sizing (affecting $C_{load, i}$).

\subsection{Temporal Independence Assumption}
To efficiently compute $\alpha_i$ without reliance on explicit input vectors, we employ Static Probability, denoted as $P(x=1)$ or simply $P$, which represents the probability that a signal $x$ is at logic `1'. Under the assumption of Temporal Independence (where the state of a signal at time $t$ is independent of its state at time $t-1$), the toggle rate (TR) is derived solely from the static probability \cite{najm1994power}:
\begin{equation}
    \mathrm{TR}_{model} = 2 \cdot P(x=1) \cdot (1 - P(x=1))
\label{eq:trmodel}
\end{equation}
This equation models the probability of a transition occurring (0$\rightarrow$1 or 1$\rightarrow$0) based on the randomness of the data. While this approximation neglects temporal correlations between clock cycles and spatial correlations (e.g., reconvergent fan-out, where the same logical signal reaches a gate along multiple paths and invalidates the independence assumption), it yields a differentiable closed-form expression. This allows us to analytically compute partial derivatives of power with respect to probability.
Crucially, for gradient-guided optimization, the model need only \emph{rank} pins correctly by activity level, not predict absolute toggle rates, for the gradients to point the optimizer in the right direction. Section~\ref{method} incorporates spatial correlations through a bit-packed simulation pass fused with the analytical model, preserving this ranking fidelity across designs with reconvergent structure.

\subsection{Power Virus and Transition-Weighted Power}
A power virus is an input (or state-transition) pattern that induces near-worst-case or worst-case switching power in a design. Power virus generation is important for (i)~worst-case power and thermal validation, (ii)~stress testing of power delivery and packaging, and (iii)~design margin verification. For combinational logic, the problem is: given primary inputs (or combinational startpoints) at two consecutive time steps, choose vectors $\mathbf{v}_0$ and $\mathbf{v}_1$ so that the resulting transition-weighted power is maximized. Formally, after propagating $\mathbf{v}_0$ and $\mathbf{v}_1$ through the netlist to obtain logic values at every pin, the transition-weighted power (the quantity we maximize) is
\begin{equation}
    P_{virus} = \sum_{i \in \mathrm{pins}} C_{load,i} \cdot \mathbb{I}\bigl( \mathrm{val}_i(\mathbf{v}_0) \neq \mathrm{val}_i(\mathbf{v}_1) \bigr)
\label{eq:pvirus}
\end{equation}
where $\mathrm{val}_i(\mathbf{v})$ is the logic value at pin $i$ under input $\mathbf{v}$, and $\mathbb{I}(\cdot)$ is the indicator function. This is the same metric used in vector-based sign-off: it counts capacitance-weighted transitions. The computation of the true power virus is an NP-hard problem.
Existing approaches, such as evolutionary search and greedy one-bit flip, rely on discrete evaluations of $P_{virus}$, rendering them computationally intractable for modern million-gate designs. Because greedy search requires $O(|\mathcal{S}|^2)$ evaluations per restart (flipping each of the $|\mathcal{S}|$ startpoints and evaluating), combinatorial heuristics completely break down at scale, effectively trapping designers in local optima beyond a few thousand inputs.
Evolutionary search scales better but still requires many function evaluations and population management, with runtimes reaching hours on industrial-scale designs.
A differentiable approximation enables gradient ascent in the continuous probability space, exploring the $|\mathcal{S}|$-dimensional landscape in $O(T \cdot |\text{pins}|)$ time for $T$ gradient steps, orders of magnitude faster than population-based or combinatorial methods.

%% file: tex/3_method.tex
\section{Propagation Methodology} \label{method}

The core of the DiffPower framework is a GPU-resident graph engine capable of performing both forward evaluation (inference) and backward differentiation (sensitivity analysis). By keeping the entire graph structure in GPU memory, we avoid the high latency of host-to-device data transfer during iterative updates. Fig.~\ref{fig:overview} shows a high-level overview of the framework.

\subsection{Graph Construction and Levelization}

We use a Tcl extraction script executed within the commercial host environment (e.g., Fusion Compiler) to extract structural connectivity, physical attributes (specifically $C_{load}$ for every net and pin), and functional definitions of the cells within a design.

We then construct a DAG representation of the circuit similar to previous works~\cite{lu2025insta, li2026deftdifferentiableautomatictest, petersen2022differentiablelogic}. Nodes represent individual pins (rather than cells), enabling fine-grained modeling of both intra-cell and inter-cell arcs. To preserve a DAG structure, the outputs of sequential elements (flip-flops, latches) are treated as pseudo-primary inputs (PPIs) with fixed probability values, while their inputs serve as pseudo-primary outputs (PPOs). Infrastructure nets, such as clock buffers, test-enable lines, and scan logic, are handled identically: they become fixed boundaries with fixed static probabilities, excluded from gradient updates while their downstream load contributions are retained.
The analysis is thus confined to the combinational logic clouds between sequential boundaries, following the same scope adopted by prior differentiable gate-level analysis frameworks~\cite{lu2025insta}, which contain the majority of optimization opportunities.

To maximize GPU parallelism, the graph is levelized: nodes are assigned levels such that a node at level $L$ depends exclusively on nodes at levels $< L$. All nodes at a given level then execute in a single parallel kernel launch.

DiffPower features an automated preprocessing pipeline (Algorithm~\ref{alg:bytecode_preprocess}) that translates Boolean expressions into GPU-executable bytecode. A lexer and recursive-descent parser (precedence: NOT $>$ AND $>$ XOR $>$ OR) build an AST, which is compiled in post-order into stack-machine opcodes: \texttt{IN}$(k)$, \texttt{NOT}, \texttt{AND}, \texttt{OR}, \texttt{XOR}, \texttt{CONST\_0/1}. The preprocessor deduplicates expressions by bytecode signature and assigns compact $logic\_id$ indices.

Common gate types (AND, OR, NAND, NOR, XOR, BUF, INV) are recognized and routed to hand-optimized CUDA kernels for slightly higher throughput. All remaining cells, regardless of complexity, are handled by the general bytecode interpreter, which requires no gate-specific code. Because all threads at a given level execute the same interpreter loop (differing only in their bytecode offset), warp divergence is minimized. This enables the representation to scale to arbitrary PDK cell libraries with $O(|V|)$ complexity per level, where $|V|$ is the number of nodes at that level.

\input{algorithms/1_bytecode_preprocess}

\subsection{Pass 1: Differentiable Forward Propagation}
DiffPower performs a differentiable algebraic propagation of static probability (detailed in Algorithm~\ref{alg:forward_prop}). The host iterates over levels; at each level two CUDA kernels execute in sequence. Kernel~1 handles wire (net) nodes by copying the driver's SP to every sink. Kernel~2 handles cell-output nodes: each thread gathers predecessor SPs via Compressed Sparse Row (CSR) indexing and dispatches to either a hand-optimized SP formula (common gates) or the general bytecode stack-machine interpreter (arbitrary cells) based on $logic\_id$. A final kernel estimates $\mathrm{TR}_{model}$ from the temporal-independence assumption (Eq.~\eqref{eq:trmodel}).
All SP operations are differentiable, enabling exact gradient computation in the backward pass (Section~\ref{backwards}). The complete forward pass runs in $O(|E|)$ time, where $|E|$ is the number of edges in the pin graph, and total switching power is computed by Eqs.~\eqref{eq:psw} and~\eqref{eq:trmodel}.
\input{algorithms/2_forward_propagation}

\subsection{Pass 2: Bit-Packed Parallel Simulation}
The temporal independence model misses spatial correlations (e.g., reconvergent fan-out). To capture these, we implement a bit-packed parallel simulation pass inspired by GATSPI~\cite{zhang2022gatspigpuacceleratedgatelevel}, computing per-pin $\mathrm{TR}_{acc}$ values that are fused with the analytical estimates in the hybrid step (Section~\ref{sec:hybrid}).
64-bit input vectors are generated via Markov chains calibrated to target SP and TR values. Startpoints sharing an upstream sequential driver receive the same bitstream to preserve correlations. The accurate toggle rate is computed as:
\begin{equation}
    \mathrm{TR}_{acc} = \frac{1}{N_{\mathrm{sim}}} \sum_{cycles} \text{PopCount}(vec_t \oplus vec_{t-1})
\end{equation}

\subsection{Hybrid Methodology} \label{sec:hybrid}
Passes~1 and~2 are fused into a single hybrid propagation pipeline. For each pin, the ratio of the simulated toggle rate to the analytical estimate defines a per-pin scaling factor $\beta_i$:
\begin{equation}
    \beta_i = \text{clip}\left( \frac{\mathrm{TR}_{acc, i}}{\mathrm{TR}_{model, i}}, \beta_{min}, \beta_{max} \right)
\end{equation}
Clipping bounds prevent gradient suppression or explosion. The clipped $\beta_i$ scales the forward activity values reported as Hybrid in Section~\ref{experiments} and, during backpropagation, the analytical gradient, so that both reflect spatial correlations captured by the simulation pass:
\begin{equation}
    \frac{\partial P_{total}}{\partial P_i} \leftarrow \beta_i \cdot \frac{\partial P_{total}}{\partial P_i}
\end{equation}
This per-pin scaling is analogous to the Straight-Through Estimator (STE) used in quantized neural networks~\cite{hubara2018quantized} and in EDA for learning-driven gate sizing~\cite{ye2025learning}. Biased gradient estimators can still converge in non-convex settings under mild conditions~\cite{yin2019understanding}, and dynamic scaling prevents vanishing gradients~\cite{le2022adaste}.
The simplicity of this estimator is intentional: $\beta_i$ requires no learned parameters and adds only a single element-wise multiply during backpropagation. The core contribution is the end-to-end hybrid pipeline that produces simulation-informed analytical gradients at GPU speed, a capability no prior power analysis tool provides.
\subsection{Gradient Backpropagation} \label{backwards}

DiffPower computes exact power gradients by traversing the levelized graph in reverse topological order (analogous to backpropagation in neural networks~\cite{lu2025insta, ji2025diffccd}). For each node $i$, the gradient of total switching power with respect to its SP is accumulated from downstream nodes and a local power term:
\begin{equation}
    \nabla_i = \frac{\partial P_{total}}{\partial P_i} = \sum_{j \in \mathrm{fanout}(i)} \nabla_j \cdot \frac{\partial P_j}{\partial P_i} + \frac{\partial P_{local}}{\partial P_i}
\end{equation}
The local term follows from differentiating the temporal-independence model (Eq.~\eqref{eq:trmodel}): $\partial P_{sw}/\partial P_i \propto C_i \cdot (1 - 2P_i)$. For common gate types, the partial derivatives $\partial P_j/\partial P_i$ are evaluated via hand-optimized backward kernels (e.g., $\partial P_{\mathrm{AND}}/\partial P_A = P_B$ for AND($A$, $B$)).

For all other cells, DiffPower employs reverse-mode AD directly over the bytecode instruction sequence (Algorithm~\ref{alg:bytecode_backward}). During the forward pass, operands of each binary operation are recorded into a per-gate tape. The backward pass traverses bytecode in reverse, applying chain-rule derivatives:
\begin{itemize}
    \item \textbf{AND}$(a,b)$: $\partial(ab)/\partial a = b$, $\partial(ab)/\partial b = a$
    \item \textbf{OR}$(a,b) = 1-(1-a)(1-b)$: $\partial/\partial a = 1-b$, $\partial/\partial b = 1-a$
    \item \textbf{XOR}$(a,b) = a+b-2ab$: $\partial/\partial a = 1-2b$, $\partial/\partial b = 1-2a$
    \item \textbf{NOT}$(x) = 1-x$: passes gradient with sign flip $(-1)$
\end{itemize}
\texttt{IN}$(k)$ accumulates its gradient into the predecessor pin via atomic addition. This yields exact gradients for arbitrary PDK cells with no gate-specific code; the backward pass has the same $O(|E|)$ complexity as the forward pass.
Gradient correctness is verified against finite-difference estimates
both on sampled root pins and via exhaustive per-pin sweeps (Section~\ref{sec:grad_val}).

\input{algorithms/3_bytecode_backward}

%% file: algorithms/1_bytecode_preprocess.tex
\begin{algorithm}[t]
\caption{PDK Bytecode Compilation (Preprocessing)}
\label{alg:bytecode_preprocess}
\begin{algorithmic}[1]
\State \textbf{Input:} \texttt{functions.csv} with rows $(ref\_name, pin\_name, func\_str)$
\State \textbf{Output:} \texttt{generated\_logic.json} --- flat bytecode arrays with per-gate metadata

\State $ops\_arr \gets [\,]$; $args\_arr \gets [\,]$
\State $offsets \gets \{\}$; $lengths \gets \{\}$; $id\_map \gets \{\}$
\State $next\_id \gets \texttt{GEN\_START}$

\For{each unique $(ref\_name, out\_pin)$ logic key $k$}
    \State $func\_str \gets$ lookup from \texttt{functions.csv}
    \Statex \hspace{2em}\textit{// Tokenize: IDENT, AND, OR, XOR, NOT, PAREN, CONST}
    \State $tokens \gets \Call{Lexer}{func\_str}$
    \Statex \hspace{2em}\textit{// Parse with precedence: NOT $>$ AND $>$ XOR $>$ OR}
    \State $ast \gets \Call{RecursiveDescentParser}{tokens}$
    \State $input\_order \gets \Call{CollectInputOrder}{ast}$
    \State $sym\_idx \gets \{name \mapsto i : (i, name) \in \text{enumerate}(input\_order)\}$
    \Statex \hspace{2em}\textit{// Post-order; ops: IN, NOT, AND, OR, XOR, CONST}
    \State $bc \gets \Call{ASTToBytecode}{ast,\; sym\_idx}$
    \Statex \hspace{2em}\textit{// Check if bytecode matches a known common gate}
    \State $known \gets \Call{MatchKnownGate}{bc,\; input\_order}$
    \If{$known \neq \texttt{None}$}
        \State $id\_map[k] \gets known$ \Comment{Use hardcoded gate ID}
    \Else
        \State $offsets[k] \gets |ops\_arr|$; $lengths[k] \gets |bc|$
        \State $ops\_arr.\text{extend}([op \text{ for } (op,\_) \in bc])$
        \State $args\_arr.\text{extend}([arg \text{ for } (\_,arg) \in bc])$
        \State $id\_map[k] \gets next\_id$; $next\_id \mathrel{+}= 1$
    \EndIf
\EndFor
\State \textbf{Export} $\{id\_map,\; ops\_arr,\; args\_arr,\; offsets,\; lengths\}$
\end{algorithmic}
\end{algorithm}

%% file: algorithms/2_forward_propagation.tex
\begin{algorithm}[t]
\caption{GPU Levelized Forward SP Propagation}
\label{alg:forward_prop}
\begin{algorithmic}[1]
\State \textbf{Input:} Levelized pin graph $G$ with $L$ levels; CSR adjacency (\textit{indptr}, \textit{indices});
\Statex \hspace{4.3em} $logic\_id[\cdot]$, compiled bytecode arrays, input permutation $\mathit{perm}[\cdot]$, startpoint SPs $P[\cdot]$
\State \textbf{Output:} Propagated SP $P[\cdot]$ and modeled TR $\hat{\alpha}[\cdot]$ for all $N$ pins

\For{$\ell = 0$ \textbf{to} $L - 1$} \Comment{Host launches two kernels per level}
    \Statex \hspace{2em}\textit{Kernel 1 --- Wire propagation (1 CUDA thread per net-sink at level $\ell$):}
    \State $\mathit{tid} \gets \texttt{blockIdx} \cdot \texttt{blockDim} + \texttt{threadIdx}$
    \State $i \gets \mathit{netFrontier}_\ell[\mathit{tid}]$ \Comment{Map thread to net-sink pin}
    \State $P[i] \gets P[\,\mathit{indices}[\mathit{indptr}[i]]\,]$ \Comment{Copy driver's SP to sink}
    \Statex \hspace{2em}\textit{Kernel 2 --- Cell evaluation (1 CUDA thread per cell-output at level $\ell$):}
    \State $i \gets \mathit{cellFrontier}_\ell[\mathit{tid}]$ \Comment{Map thread to cell-output pin}
    \State Gather $a_k \gets P[\,\mathit{indices}[\mathit{indptr}[i]+k]\,]$ for $k = 0,\ldots,n_i{-}1$
    \If{$logic\_id[i] < \texttt{GEN\_START}$}
        \Statex \hspace{4em}\textit{// e.g.\ AND: $a_0 \cdot a_1$; NAND: $1 - a_0 \cdot a_1$}
        \State $P[i] \gets \Call{HandOptimizedSP}{logic\_id[i],\; a_0, \ldots, a_{n_i-1}}$
    \Else
        \State Permute via $\mathit{perm}[i]$; run bytecode interpreter
        \State $P[i] \gets \mathrm{clip}(\text{stack top},\; 0,\; 1)$
    \EndIf
\EndFor
\Statex \textit{Kernel 3 --- TR estimation ($N$ threads, all pins in parallel):}
\State $\hat{\alpha}[i] \gets 2 \cdot P[i] \cdot (1 - P[i])$ for all pins $i$ \Comment{Eq.~\eqref{eq:trmodel}}
\end{algorithmic}
\end{algorithm}

%% file: algorithms/3_bytecode_backward.tex
\begin{algorithm}
\caption{Bytecode Backward Pass (Reverse-Mode AD)}
\label{alg:bytecode_backward}
\begin{algorithmic}[1]
\State \textbf{Input:} Bytecode $bc$ of length $L$; forward tape $\mathit{tape}_a[0..L{-}1]$, $\mathit{tape}_b[0..L{-}1]$;
\Statex \hspace{4.3em} upstream gradient $g_{out}$; number of gate inputs $n$
\State \textbf{Output:} Per-input gradients $\nabla_{in}[0..n{-}1]$

\State $\nabla_{in}[i] \gets 0$ for $i \in [0, n)$
\State Initialize backward grad-stack: $\mathit{gstk} \gets [g_{out}]$

\For{$i \gets L - 1$ \textbf{downto} $0$}
    \State $op, arg \gets bc[i]$
    \If{$op = \texttt{IN}$}
        \State $g \gets \mathit{gstk}.\text{pop}()$
        \State $\nabla_{in}[arg] \mathrel{+}= g$
    \ElsIf{$op = \texttt{NOT}$}
        \State $g \gets \mathit{gstk}.\text{pop}()$
        \State $\mathit{gstk}.\text{push}(-g)$ \Comment{$\partial(1-x)/\partial x = -1$}
    \ElsIf{$op = \texttt{AND}$}
        \State $g \gets \mathit{gstk}.\text{pop}()$; $a \gets \mathit{tape}_a[i]$; $b \gets \mathit{tape}_b[i]$
        \State $\mathit{gstk}.\text{push}(g \cdot b)$ \Comment{$\partial(ab)/\partial a = b$}
        \State $\mathit{gstk}.\text{push}(g \cdot a)$ \Comment{$\partial(ab)/\partial b = a$}
    \ElsIf{$op = \texttt{OR}$}
        \State $g \gets \mathit{gstk}.\text{pop}()$; $a \gets \mathit{tape}_a[i]$; $b \gets \mathit{tape}_b[i]$
        \State $\mathit{gstk}.\text{push}(g \cdot (1-b))$ \Comment{$\partial/\partial a = 1{-}b$}
        \State $\mathit{gstk}.\text{push}(g \cdot (1-a))$ \Comment{$\partial/\partial b = 1{-}a$}
    \ElsIf{$op = \texttt{XOR}$}
        \State $g \gets \mathit{gstk}.\text{pop}()$; $a \gets \mathit{tape}_a[i]$; $b \gets \mathit{tape}_b[i]$
        \State $\mathit{gstk}.\text{push}(g \cdot (1-2b))$ \Comment{$\partial/\partial a = 1{-}2b$}
        \State $\mathit{gstk}.\text{push}(g \cdot (1-2a))$ \Comment{$\partial/\partial b = 1{-}2a$}
    \ElsIf{$op \in \{\texttt{CONST\_0},\, \texttt{CONST\_1}\}$}
        \State $\mathit{gstk}.\text{pop}()$ \Comment{Constants have no input dependence}
    \EndIf
\EndFor
\end{algorithmic}
\end{algorithm}

%% file: tex/4_optimizations.tex
\section{Gradient-Based Optimization Frameworks} \label{optimizations}

The backward pass of Section~\ref{method} yields what we term \emph{power gradients}: the derivatives $\partial P_{\mathrm{total}}/\partial P_{\mathrm{sw},i}$ of total switching power with respect to each pin's activity. These are obtained from the SP gradients $\partial P_{total}/\partial P_i$ computed in Section~\ref{backwards} via the chain rule through Eq.~\eqref{eq:trmodel}.
Unlike local metrics that consider only a cell's own power consumption, power gradients capture downstream effects, revealing how a change at one pin propagates through fan-out and affects power throughout the design.
This global sensitivity information is unavailable to conventional power analysis tools and can, in principle, inform any power-aware optimization that benefits from knowing how local activity changes propagate globally. We demonstrate this with two representative applications: gradient-ascent power virus generation and gradient-weighted cell sizing for power reduction.

\subsection{Gradient-Weighted Timing-Aware Cell Sizing}
Cell sizing is inherently a discrete optimization problem in which the designer selects among library-specific alternatives (e.g., AND2\_X1, AND2\_X2, AND2\_X4) that trade off drive strength (and hence capacitance) against timing margin.
Standard approaches rank cells by local capacitance or slack; we introduce a \emph{gradient-weighted ranking metric} that combines global sensitivity with local power to maximize the power reduction achieved per cell resized.

We observe that the most effective downsizing candidates exhibit both high sensitivity (where a small local change produces a large global power reduction) and high existing local power consumption.
Formally, we define the ranking score $S_i$ for cell instance $i$ as:
\begin{equation}
    S_i = \left| \frac{\partial P_{total}}{\partial P_{sw, i}} \right| \times P_{sw, i}
\label{eq:score}
\end{equation}

The gradient term $|\partial P_{total}/\partial P_{sw,i}|$ captures the \emph{global} impact of modifying cell $i$: downsizing a cell that drives a high-fan-out net yields a large gradient because the capacitance change propagates to many downstream pins, each contributing to total power.
The local power term $P_{sw,i}$ ensures that the metric focuses on cells where there is substantial switching activity to reduce. A cell with a large gradient but near-zero activity offers no practical savings.
The product acts as a first-order estimate of the achievable power reduction from resizing cell $i$, analogous to the gain metric used in timing-driven sizers~\cite{lu2025lego, pham2025agd} but targeting switching power rather than delay.

The sizing procedure operates iteratively.
At each iteration, cells not on a critical timing path are ranked by $S_i$ in descending order.
For the top-ranked cell, all library-specific legal substitutions are identified using the \texttt{get\_alternative\_lib\_cells} command in the commercial environment.
Each candidate is evaluated: the substitution is accepted if it reduces switching power without creating a timing violation. Slack is queried directly from the commercial timer after each trial substitution, and any candidate introducing negative slack on any path through the cell is rejected, so accepted swaps remain sign-off-consistent.
Because the global gradient changes as cells are resized (a phenomenon known as \emph{gradient staleness}), the ranking score becomes gradually less accurate over successive iterations.
In practice, we process cells in small batches and periodically re-invoke DiffPower's forward and backward passes to refresh the gradient, maintaining ranking fidelity at trivial computational cost. Re-propagation takes $\sim$27\,ms on the small design and $\sim$33\,ms on the medium design, multiple orders of magnitude faster than existing methods.

\input{algorithms/4_power_virus}
\subsection{Power Virus Generation via Gradient Ascent} \label{sec:power_virus_method}
The true $P_{virus}$ objective (Eq.~\eqref{eq:pvirus}) is discrete and non-differentiable, counting capacitance-weighted bit-flips.
We relax the problem by treating startpoint SPs as continuous variables in $[0,1]$ and maximizing the differentiable proxy:
\begin{equation}
    P_{proxy} = \sum_{i \in \mathrm{pins}} C_i \cdot 2\,P_i\,(1 - P_i)
\label{eq:pproxy}
\end{equation}
which equals the expected transition probability at each pin under the temporal-independence model.
Starting from a uniform initialization ($P_i{=}0.5$ for all startpoints), each gradient ascent step computes $\nabla_{P_i} P_{proxy}$ via one forward/backward pass and updates $P_i \leftarrow \mathrm{clip}(P_i + \eta \cdot \nabla_i,\, 0,\, 1)$.
After $T$ steps (typically $T{=}100$), we discretize by thresholding: $v_{0,i}{=}\mathbb{I}(P_i \ge 0.5)$ and $v_{1,i}{=}1{-}v_{0,i}$, producing an input pair that maximizes transitions at the startpoints.
Two deterministic forward passes evaluate the true $P_{virus}$ on the discrete vectors.
The entire procedure runs in $O(T \cdot |\text{pins}|)$ time and requires no population management, fitness evaluation budget, or discrete search heuristics.
A one-time bit-packed run can optionally compute $\beta$ factors to scale gradients during ascent, improving the search direction when the temporal-independence model misestimates toggle rates. Algorithm~\ref{alg:power_virus} summarizes the full procedure.

%% file: algorithms/4_power_virus.tex
\begin{algorithm}[t]
\caption{DiffPower Gradient-Ascent Power Virus Search}
\label{alg:power_virus}
\begin{algorithmic}[1]
\State \textbf{Input:} Graph $G$, startpoints $\mathcal{S}$, steps $T$, rate $\eta$, optional $\beta$-scaling
\State \textbf{Output:} Vectors $\mathbf{v}_0, \mathbf{v}_1$ maximizing $P_{virus}$
\State \textbf{Notation:} $P_i$ = SP at startpoint $i$; $\nabla_i = \partial P_{total} / \partial P_i$
\State Initialize $P_i \gets 0.5$ for all $i \in \mathcal{S}$
\If{$\beta$-scaling enabled}
    \Statex \hspace{2em}\textit{// One-time bit-packed simulation for spatial correlations}
    \State Compute $\beta_j \gets \mathrm{TR}_{acc,j} / \mathrm{TR}_{model,j}$ per pin $j$
    \State Store $\beta$ for use in backpropagation
\EndIf
\For{$t = 1$ \textbf{to} $T$}
    \Statex \hspace{2em}\textit{// Map startpoint SPs to full pin array, propagate, backprop}
    \State $\mathbf{a} \gets \Call{BuildSPArray}{P_1,\ldots,P_{|\mathcal{S}|}}$
    \State $\Call{ForwardPropagationGPU}{G, \mathbf{a}}$
    \State $\nabla \gets \Call{BackwardPropTotalPower}{G}$ \Comment{Scaled by $\beta$}
    \For{each startpoint $i \in \mathcal{S}$}
        \State $P_i \gets P_i + \eta \cdot \nabla_i$ \Comment{Ascent step}
        \State $P_i \gets \mathrm{clip}(P_i, 0, 1)$
    \EndFor
\EndFor
\Statex \textit{// Discretize continuous SPs to binary vectors}
\State $v_{0,i} \gets \mathbb{I}(P_i \ge 0.5)$,\; $v_{1,i} \gets 1 - v_{0,i}$ for all $i \in \mathcal{S}$
\Statex \textit{// Evaluate true $P_{virus}$ on discrete vectors via bitwise simulation}
\State $P_{virus} \gets \Call{EvaluatePowerFromVectors}{G, \mathbf{v}_0, \mathbf{v}_1}$
\State \Return $\mathbf{v}_0, \mathbf{v}_1, P_{virus}$
\end{algorithmic}
\end{algorithm}

%% file: tex/5_experiment.tex
\section{Experimental Results} \label{experiments}

\subsection{Experimental Setup}
Experiments were conducted on a single NVIDIA A100 GPU (40\,GB VRAM) and an AMD EPYC 7F52 16-Core host. The accuracy baseline is commercial switching-activity propagation run through Synopsys Fusion Compiler or Design Compiler.
We evaluate DiffPower on ten designs. Three are proprietary combinational netlists implemented in different technology nodes, extracted with Fusion Compiler: a small design with 13K cells, a medium design with 117K cells, and a larger deep-learning accelerator with 652K cells, included deliberately to stress-test the combinational-boundary assumption at scale. The remaining seven are IWLS\,2005 open-source benchmarks~\cite{iwls2005} synthesized with Design Compiler.
All comparisons are scoped to combinational switching activity, consistent with prior differentiable gate-level analysis frameworks~\cite{lu2025insta, li2026deftdifferentiableautomatictest}: sequential elements are treated as analysis boundaries, with their output SP and TR taken from the design database.
The commercial reference tool is run with \texttt{propagate\_switching\_activity} from the same register-output starting conditions, giving a like-for-like comparison of combinational propagation accuracy.

\input{tables/2_prop_times}

\begin{figure}[t]
\centering

\begin{subfigure}{0.98\linewidth}
\centering
\includegraphics[width=.95\linewidth]{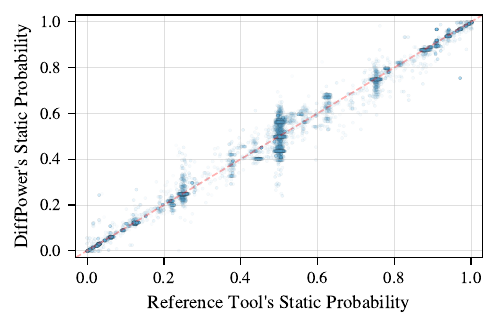}
\caption{Hybrid static probability (SP) computed by DiffPower vs. commercial reference for all pins in the small design ($r = 0.993$).}
\label{fig:modeled_afp}
\end{subfigure}

\vspace{2pt}

\begin{subfigure}{0.98\linewidth}
\centering
\includegraphics[width=.95\linewidth]{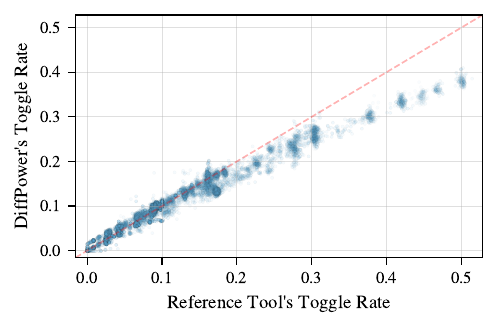}
\caption{Hybrid toggle rate (TR) computed by DiffPower vs. commercial reference for all pins in the small design ($r = 0.982$).}
\label{fig:corrected_tr_100}
\end{subfigure}

\caption{Per-pin activity correlation between DiffPower and a commercial reference tool on the small design (53K pins).}
\label{fig:activity_correlation}
\end{figure}

\input{tables/3_prop_results}

\subsection{Propagation Runtime \& Accuracy}

We evaluate three propagation modes: (1)~\textbf{Modeled} algebraic SP propagation under the temporal-independence assumption, (2)~\textbf{Bitwise} 64-bit parallel simulation capturing spatial correlations, and (3)~\textbf{Hybrid:} the full pipeline fusing analytical SP with per-pin $\beta$-scaling from bitwise TR.

Table~\ref{tab:prop_times} reports GPU kernel execution times, single-threaded CPU DiffPower times, and commercial reference tool times across the three proprietary designs.
DiffPower completes forward propagation and backpropagation on a 652K-cell design in under 60\,ms, approximately $1{,}002\times$ faster than single-threaded CPU propagation and three to four orders of magnitude faster than the commercial reference tool.

Table~\ref{tab:prop_results} presents per-pin SP and TR correlation against the commercial reference across all ten designs.
DiffPower's hybrid pipeline (100 Monte Carlo iterations, per-pin $\beta$-clipping with $\beta_{min}=0.1$, $\beta_{max}=10$) achieves a median SP correlation of $r{=}0.988$ and median TR correlation of $r{=}0.96$ across all ten designs.
On the nine well-conditioned designs, SP~$r \geq 0.98$ and TR~$r$ ranges from $0.937$ to $0.995$, indicating that per-pin activity rankings closely match the commercial reference.
For comparison, the analytical-only (Modeled) baseline achieves high SP correlation (median $r{=}0.997$) but lower TR correlation (median $r{=}0.86$), reflecting the spatial-correlation gap that the bit-packed simulation pass addresses.
The large proprietary extraction (652K cells) establishes the current operational boundary, achieving TR~$r=0.84$, a limit of the combinational-boundary approximation analyzed in Section~\ref{discussion}.

Figs.~\ref{fig:modeled_afp} and~\ref{fig:corrected_tr_100} show per-pin SP and TR scatter plots for DiffPower's hybrid propagation on the small design.
SP values maintain tight alignment with the commercial reference ($r = 0.993$, Fig.~\ref{fig:modeled_afp}), and TR values achieve $r = 0.982$ (Fig.~\ref{fig:corrected_tr_100}), confirming that per-pin activity rankings closely match the commercial tool and that the resulting power gradients are well-directed for optimization.

Table~\ref{tab:prop_results} also shows the memory footprint of each design to store the DAG and propagation value buffers. This scales linearly with pin count, from 35.8~KB on the smallest benchmark to 213.5~MB on the largest design, all fitting comfortably within a single GPU's VRAM.

\subsection{Gradient Validation} \label{sec:grad_val}
We validate gradient correctness at two levels: pointwise accuracy and large-scale rank agreement.

\textbf{Pointwise finite-difference validation.} For each of 32 randomly sampled root pins, we perturb the root SP by $\varepsilon=10^{-4}$ and re-propagate to obtain a one-sided finite-difference gradient estimate.
On five designs (\texttt{sasc}, \texttt{spi}, \texttt{pci}, \texttt{aes\_cipher}, \texttt{medium}), all 160 tested pins achieve 100\% pass rate with mean relative error below $10^{-4}$ and maximum relative error below $2\times 10^{-4}$, confirming that reverse-mode AD through the bytecode interpreter produces exact derivatives of the forward SP model.
Extending this validation to full-design scale, we run exhaustive finite-difference sweeps over all root-driven pins for \texttt{aes\_core} and \texttt{pci}. The median relative error is $2.71\times 10^{-4}$ and $1.50\times 10^{-4}$, respectively; the median is reported because a heavy tail of near-zero gradient magnitudes inflates the relative-error denominator. Top-100/500/1{,}000 rank overlaps against the brute-force gradients are $100/499/1{,}000$ on \texttt{aes\_core} and $99/486/990$ on \texttt{pci}, confirming that the analytical backward pass reproduces numerical sensitivity rankings almost exactly across entire designs.

\textbf{Gradient rank correlation.} Because the downstream applications rely on gradient-ranked pin ordering rather than absolute gradient magnitudes, rank agreement with ground truth is the critical validation metric. To assess whether analytical gradients identify the same high-sensitivity startpoints as brute-force numerical differentiation, we run a large-scale finite-difference sweep on the \textbf{medium} industrial design (117K cells, $50{,}222$ total roots).
We evaluate a stratified subset of $15{,}000$ roots, biased toward high-$|\nabla|$ roots to focus on the optimizer-relevant portion of the distribution.
The GPU analytical gradient completes in 68\,s (dominated by graph creation overhead followed by one forward/backward pass); the CPU finite-difference sweep ($\varepsilon{=}10^{-4}$, one perturbation per root) requires $61{,}668$\,s ($\sim$17\,hours), a $904\times$ speedup.
The two gradient vectors achieve Pearson~$r{\approx}1.0$ and top-$k$ Jaccard$\,{=}\,1.0$ for all $k \leq 500$, confirming \textit{near-perfect} gradient agreement at the scale most relevant to production optimization.

\subsection{Optimization Quality}
We evaluate the efficacy of DiffPower's guided cell sizing by comparing the power reduction from iterative cell sizing against existing methodologies.

\begin{figure}[!t]
\centering

\begin{subfigure}{0.98\linewidth}
\centering
\includegraphics[width=.95\linewidth]{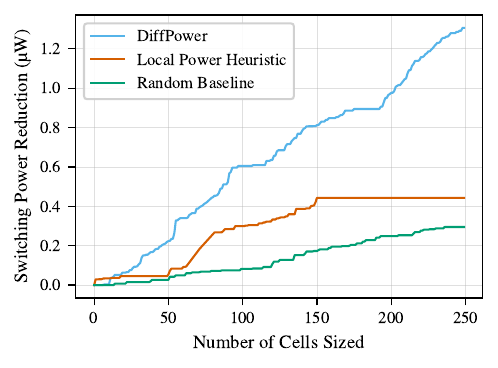}
\caption{Small design (13K cells). After 250 iterations, DiffPower achieves a reduction of $1.31\,\mu\text{W}$, outperforming Local Power ($0.44\,\mu\text{W}$) by $2.98\times$ and Random ($0.30\,\mu\text{W}$) by $4.4\times$.}
\label{fig:sizing_results}
\end{subfigure}

\begin{subfigure}{0.98\linewidth}
\centering
\includegraphics[width=.95\linewidth]{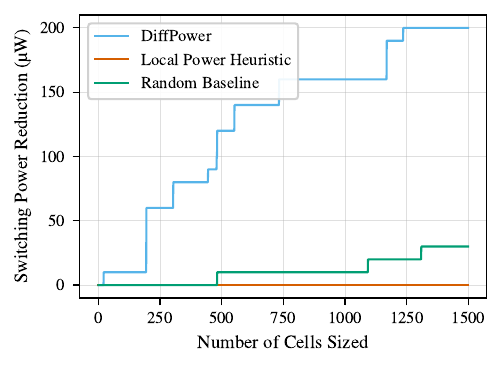}
\caption{Medium design (117K cells). After 1{,}500 iterations, DiffPower achieves a reduction of $200\,\mu\text{W}$ while Random reaches only $30\,\mu\text{W}$ and Local Power effectively zero}.
\label{fig:sizing_medium}
\end{subfigure}

\caption{Cumulative switching power reduction during iterative cell sizing. Cells are processed in rank order; three ordering strategies are compared: random selection, local power ($\sum C \cdot \text{TR}$), and DiffPower's gradient-weighted metric ($|\nabla P_{total}| \times P_{sw,i}$).}
\label{fig:sizing}
\end{figure}

Fig.~\ref{fig:sizing} shows the cumulative power reduction under the iterative cell sizing scheme on the small (a) and medium (b) industrial designs.
Three cell-ordering strategies are compared: random selection, local power ($\sum C \cdot \text{TR}$), and DiffPower's power-gradient-weighted metric ($|\nabla P_{total}| \times P_{sw,i}$).
On the small design, the gradient-weighted metric outperforms local power by $2.98\times$ after 250 cells resized.
On the medium design, the advantage is even more pronounced: after 1{,}500 iterations, DiffPower achieves $200\,\mu\text{W}$ cumulative reduction while random achieves only $30\,\mu\text{W}$ and local power achieves zero.
Power gradients capture both local power and global fan-out effects, explaining why the advantage grows with design complexity. Deeper logic cones amplify the value of global sensitivity information: local heuristics that ignore downstream impact plateau, while gradient-weighted sizing continues to find productive swaps.
These curves (especially Fig.~\ref{fig:sizing_medium}) exhibit flat regions where no power gain is realized; these occur when the top-ranked cell has no legal replacement that improves power without introducing negative slack. The sizer skips such cells and advances to the next candidate, resuming reduction once a viable swap is found. This is especially notable in the local power heuristic, which struggles to identify any timing-feasible downsizing alternatives.

\subsection{Power Virus Generation} \label{sec:exp_power_virus}
We compare DiffPower's gradient-ascent power virus against an evolutionary search baseline on ten designs (seven IWLS benchmarks plus the three industrial designs).
Both methods use the same cap-weighted transition metric $P_{virus}$ (Eq.~\eqref{eq:pvirus}). The gradient method runs $T{=}100$ forward and backward passes in $O(T \cdot |\text{pins}|)$ time. Evolutionary search is given a 1-hour wall-clock budget with population fitness evaluations parallelized on DiffPower's GPU propagation engine, giving the evolutionary baseline every computational advantage.
We focus on the evolutionary comparison because it is the only baseline that scales to large designs; greedy one-bit flip ($O(|\mathcal{S}|^2)$ per restart) is intractable for designs with even thousands of cells.

Table~\ref{tab:power_virus_quality} reports per-design results.
Gradient ascent outperforms evolutionary search on all ten designs, achieving up to $+113\%$ on \texttt{aes\_core} and $+96\%$ on the medium design (117K cells), a $1.96\times$ improvement.
With $\beta$-calibration ablated on \texttt{aes\_core} and \texttt{pci}, purely analytical gradients still beat the evolutionary baseline ($2{,}318.3$ vs.\ $1{,}110.2$; $733.4$ vs.\ $481.4$), validating the continuous relaxation; $\beta$-calibration contributes a further 2--5\%.

The runtime advantage is decisive.
On the medium design, gradient ascent completes in 74\,s and achieves a $1.96\times$ higher $P_{virus}$ score than evolutionary search given a 1-hour budget.
On the large proprietary design (652K cells), gradient ascent completes in 615\,s and outperforms evolutionary search, while greedy methods ($O(|\mathcal{S}|^2)$) are entirely infeasible at this scale.

\input{tables/4_power_virus_results}

%% file: tables/2_prop_times.tex
\begin{table}[t]
\centering
\caption{Propagation runtime on the three proprietary designs.
GPU columns are per-pass kernel times; CPU is DiffPower algebraic propagation running on CPU instead; speedup is algebraic vs.\ algebraic; Ref.\ is the commercial reference tool.
Design sizes: small (13K comb.\ cells), medium (117K), large (652K).}
\label{tab:prop_times}
\setlength{\tabcolsep}{3pt}
\begin{tabular}{@{}lcccccc@{}}
\toprule
 & \multicolumn{3}{c}{GPU Kernel (s)} & CPU & Ref. & GPU vs. \\
\cmidrule(lr){2-4}
Design & Alg. & Bitwise & Backward & (s) & (s) & CPU \\
\midrule
Small  & 0.0135 & 0.0124 & 0.0132 & 0.416  & 4.09   & 31$\times$ \\
Medium & 0.0161 & 0.0160 & 0.0165 & 4.546  & 14.15  & 282$\times$ \\
Large  & 0.0289 & 0.0269 & 0.0242 & 28.96  & 224.38 & 1{,}002$\times$ \\
\bottomrule
\end{tabular}
\end{table}

%% file: tables/3_prop_results.tex
\begin{table*}[t]
\centering
\caption{Propagation accuracy of the analytical (Modeled) and full-pipeline (Hybrid) methods against
the commercial reference tool. Hybrid results use 100-iteration Monte Carlo $+$ per-pin $\beta$-scaling.
Designs in the upper group are proprietary; the lower group is open-source benchmarks. Memory footprint is the amount of GPU memory used to store the graph and buffers used in propagation.}
\label{tab:prop_results}
\setlength{\tabcolsep}{5pt}
\begin{tabular}{@{}lrrcccccc@{}}
\toprule
& & &  & & \multicolumn{2}{c}{Modeled} & \multicolumn{2}{c}{Hybrid}\\
\cmidrule(lr){6-7}\cmidrule(lr){8-9}
Design & Pins & Cells & Levels  & Memory Usage & SP $r$ & TR $r$ & SP $r$ & TR $r$\\
\midrule
\multicolumn{9}{l}{\textit{Proprietary designs}} \\
\quad small testcase  & 53K  & 13K  & 70  & 3.8 MB   & 0.950  &  0.641  & 0.993  & 0.982 \\
\quad medium testcase & 453K & 117K & 84  & 32.8 MB  & 0.921  &  0.715  & 0.981  & 0.937 \\
\quad large testcase  & 2.6M & 652K & 104 & 213.5 MB & 0.970  &  0.831  & 0.969  & 0.835 \\
\midrule
\multicolumn{9}{l}{\textit{IWLS 2005 open-source benchmarks}} \\
\quad sasc        &  455   &  173  & 23 & 35.8 KB& 1.000  & 0.987  & 0.999 & 0.982  \\
\quad spi         &  1.1K  &  431  & 18 & 90.0 KB  & 0.997    & 0.992  & 0.995   & 0.961  \\
\quad i2c         &  1.8K  &  486  & 56 & 138 KB & 0.999  & 0.928  & 0.983 & 0.949  \\
\quad aes\_cipher &  4.7K  &  1.7K & 13 & 362 KB & 1.000    & 0.997  & 0.998 & 0.995  \\
\quad des         & 10.4K  &  3K   & 60 & 823 KB & 0.997  & 0.817  & 0.983 & 0.963  \\
\quad pci         & 33.4K  &  8.7K & 50 & 2.4 MB & 0.998  & 0.884  & 0.980 & 0.954  \\
\quad aes\_core   & 46K    & 13K   & 52 & 3.7 MB & 0.997  & 0.476  & 0.994 & 0.955  \\
\bottomrule
\end{tabular}
\end{table*}

%% file: tables/4_power_virus_results.tex
\begin{table}[t]
\centering
\caption{Power virus quality and runtime: gradient ascent vs.\ evolutionary search.
Both methods use the same cap-weighted transition metric $P_{virus}$ (Eq.~\eqref{eq:pvirus}).
Gradient ascent uses $T{=}100$ forward/backward passes; evolutionary search has a 1-hour budget and population evaluations are done in parallel on our GPU propagation engine.}
\label{tab:power_virus_quality}
\setlength{\tabcolsep}{3pt}
\begin{tabular}{@{}lrrrrc@{}}
\toprule
 & & \multicolumn{2}{c}{$P_{virus}$ Score} & & \\
\cmidrule(lr){3-4}
Design & Cells & Gradient & Evol. & Gap (\%) & DiffPower Time \\
\midrule
sasc          &    173 &   \textbf{7.1} &     6.7 &  $+6$ & $<$1\,s \\
spi           &    431 &   \textbf{49.1} &  32.4 & $+52$ & $<$1\,s \\
i2c           &    486 &  \textbf{36.9} &    28.8 & $+28$ & $<$1\,s \\
aes\_cipher   &  1.7K & \textbf{161.6} &   146.4 & $+10$ & $<$1\,s \\
des           &  3K &  \textbf{248.1} & 235.7 & $+5$ &   2\,s \\
pci           &  8.7K & \textbf{768.5} &   481.4 & $+60$ &   5\,s \\
aes\_core     & 13K & \textbf{2,367.2} & 1,110.2 & $+113$ &  16\,s \\
\midrule
small         & 13K & \textbf{222.7} &   182.3 & $+22$ &  20\,s \\
medium & 117K & \textbf{2,286.1} & 1,163.9 & $+96$ & 74\,s \\
large & 652K & $\mathbf{3.789{\times}10^6}$ & $3.788{\times}10^6$ & $+0.03$ & 615\,s \\
\bottomrule
\end{tabular}
\end{table}

%% file: tex/6_discussion.tex
\vspace{-0.05cm}
\section{Discussion} \label{discussion}

\textbf{Why Correlation Matters for Optimization.}
For power-gradient-guided optimization, per-pin \emph{correlation} is the metric that matters most, more so than aggregate power accuracy.
If the model correctly \emph{ranks} which pins are high-activity versus low-activity, the analytical gradients point in the right direction, and downstream decisions (cell sizing, power virus search) are well-prioritized.
A model with perfect correlation but a systematic $2\times$ scale bias would still produce correct gradient \emph{directions}, and the optimizer would simply adjust its step size.
Conversely, a model with low correlation but coincidentally low aggregate error (due to over/under-estimation cancellation) would mislead the optimizer by pointing gradients toward the wrong pins.
This is confirmed by the gradient rank experiment: on the medium design (117K cells), the top-500 analytically ranked pins match the top-500 numerically ranked pins exactly (Jaccard$\,{=}\,1.0$, Pearson~$r{\approx}1.0$).
DiffPower's hybrid pipeline achieves TR~$r \geq 0.937$ on nine of ten designs, indicating that pin rankings are highly reliable for optimization.
The large design (TR $r=0.84$) is the remaining accuracy frontier. Its degradation arises from two interacting factors: the combinational-boundary treatment cuts deep sequential feedback into wide, irregular logic cones, and within these cones heavy reconvergence violates the temporal-independence assumption faster than per-pin $\beta$-calibration can correct. Both are scoping constraints shared by all current differentiable gate-level frameworks~\cite{lu2025insta, li2026deftdifferentiableautomatictest} and analytical SP methods, not deficiencies of the gradient framework itself. Notably, even on this design SP correlation remains strong ($r=0.969$), confirming that the analytical propagation scales correctly.
In practice, power optimization is typically applied at the block or partition level (10K–200K cells), a regime where DiffPower achieves TR $r \geq 0.937$. Even at the 652K-cell scale, gradient ascent produces power virus scores competitive with evolutionary search, confirming that gradients remain directionally useful.

\textbf{Power Virus: Scalability and Runtime.}
Gradient ascent outperforms evolutionary search on all evaluated designs (Table~\ref{tab:power_virus_quality}), even when the evolutionary baseline is given a 1-hour budget with population fitness evaluations parallelized on DiffPower's own GPU engine.
The core advantage is computational: gradient ascent scales as $O(T \cdot |\text{pins}|)$ with fixed $T{=}100$ steps, completing in 74\,s on the medium design (orders of magnitude faster than evolutionary search) while achieving a $1.96\times$ higher $P_{virus}$ score.
The consistent advantage across design scales confirms that gradient-directed continuous relaxation is consistently more effective than population-based discrete search for this objective. Future hybrid approaches combining gradient initialization with discrete local refinement could further improve quality on the smallest designs where the gap is generally narrowest.

\textbf{Limitations.}
As noted above, restricting the analysis to combinational logic between sequential boundaries, the scope shared by prior differentiable gate-level frameworks~\cite{lu2025insta, li2026deftdifferentiableautomatictest}, limits accuracy on deeply sequential designs; sequential power, clock network power, and multi-voltage domain modeling remain open.
Infrastructure nets (clock buffers, test-mode signals), though preserved as fixed boundaries with load contributions retained (Section~\ref{method}), are not yet modeled or optimized explicitly; doing so remains a practical prerequisite for full-chip deployment.

%% file: tex/7_conclusion.tex
\vspace{-0.1cm}
\section{Conclusion} \label{conclusion}
\vspace{-0.1cm}

In this paper, we introduced DiffPower, a GPU-accelerated fully differentiable framework that resolves the historical trade-off between speed and accuracy in switching power analysis. By combining a PDK-agnostic bytecode representation with a hybrid propagation pipeline that fuses analytical modeling and bit-packed simulation, DiffPower delivers high-fidelity toggle-rate correlation (median $r{=}0.96$) and up to $1{,}002\times$ speedup over CPU propagation, completing forward and backward passes on a 652K-cell industrial design in under 60\,ms. This enables analytical gradient computation on industrial designs at scales exceeding 650K cells.
The key insight of this work is that the resulting \emph{power gradients} translate directly into downstream optimization value. Gradient-weighted cell sizing consistently outperforms local-power heuristics. For power virus generation, gradient ascent outperforms evolutionary search on all ten evaluated designs while reducing runtime from hours to seconds.
Key open challenges for future work include the accurate analysis of sequential-heavy designs, explicit modeling of infrastructure nets, and extending the framework to sequential power and multi-voltage domains. The underlying architecture, a levelized GPU graph with bytecode-driven AD, is not specific to combinational switching power and can serve as a foundation for these extensions.